\documentclass[11pt]{amsart}
\usepackage{amsmath}
\usepackage{slashed}
\usepackage{graphicx}

\theoremstyle{plain}
\newtheorem{theorem}{Theorem}[section]

\theoremstyle{definition}
\newtheorem{remark}[theorem]{Remark}

\numberwithin{equation}{section}

\begin{document}

\title[The Conformal Flow of Metrics and the General Penrose Inequality] {The Conformal Flow of Metrics and the General Penrose Inequality}

\author[Han]{Qing Han}
\address{Department of Mathematics\\
University of Notre Dame\\
Notre Dame, IN 46556} \email{qhan@nd.edu}

\address{Beijing International Center for Mathematical Research\\
Peking University\\
Beijing, 100871, China} \email{qhan@math.pku.edu.cn}

\author[Khuri]{Marcus Khuri}
\address{Department of Mathematics\\
Stony Brook University\\ Stony Brook, NY 11794}
\email{khuri@math.sunysb.edu}
\thanks{The first author acknowledges the support of NSF
Grant DMS-1105321. The second author acknowledges the support of
NSF Grant DMS-1708798.}

\begin{abstract}
The conformal flow of metrics \cite{Bray} has been used to successfully establish a special case of the Penrose inequality, which yields a lower bound for the total mass of a spacetime in terms of horizon area. Here we show how to adapt the conformal flow of metrics, so that it may be applied to the Penrose inequality for general initial data sets of the Einstein equations. The Penrose conjecture without the assumption of time symmetry is then reduced to solving a system of PDE with desirable properties.
\end{abstract}
\maketitle

\section{Introduction}
\label{sec1} \setcounter{equation}{0}
\setcounter{section}{1}

Let $(M,g,k)$ be an initial data set for the Einstein equations with a single asymptotically flat end. This triple consists of a 3-manifold $M$, on which a Riemannian metric $g$ and symmetric 2-tensor $k$ are defined and satisfy the constraint equations
\begin{equation}\label{0.1}
16\pi\mu = R+(Trk)^{2}-|k|^{2},\quad\quad\quad
8\pi J = \operatorname{div}(k+(Trk)g).
\end{equation}
The quantities $\mu$ and $J$ represent the energy and momentum densities of
the matter fields, respectively, whereas $R$ denotes the scalar curvature
of $g$. The dominant energy condition will be assumed $\mu\geq|J|$. This asserts that 
all measured energy densities are nonnegative, and implies that
matter cannot travel faster than the speed of light.

Null expansions measure the strength of the gravitational field
around a hypersurface $S\subset M$ and are given by
\begin{equation}\label{0.2}
\theta_{\pm}:=H_{S}\pm Tr_{S}k,
\end{equation}
where $H_{S}$ denotes the mean curvature with respect to the unit
normal pointing towards spatial infinity. More precisely, $\theta_{\pm}$
may be interpreted as the rate at which the area of a shell of light is changing
as it moves away from the surface in the outward future/past direction ($+$/$-$). 
A future or past trapped surface is defined by the inequality $\theta_{+}< 0$ or $\theta_{-}< 0$, respectively, and according to the above interpretation such a surface
lies in a strong gravitation field. If $\theta_{+}=0$ or $\theta_{-}=0$, then $S$ is referred to as a future or past apparent horizon. These surfaces naturally arise
as boundaries of future or past trapped regions. An apparent horizon will be referred to as outermost if it is not enclosed by any other apparent horizon. According to \cite{Galloway,GallowaySchoen} each component of an outermost apparent horizon must have spherical topology. These surfaces replace the role of event horizons in the formulation of the Penrose inequality. In particular, under the above hypotheses on the initial data, the standard version of the Penrose conjecture \cite{Mars} asserts that
\begin{equation}\label{0.3}
M_{ADM}\geq\sqrt{\frac{A_{min}^{\pm}}{16\pi}}
\end{equation}
where $M_{ADM}$ is the ADM mass and $A_{min}^{+}$ ($A_{min}^{-}$) is the minimum area required to enclose the outermost 
future (past) apparent horizon. The Riemannian version of this inequality in the time symmetric case, when $k=0$, along with the corresponding rigidity statement
in the case of equality, have been established. For a single component horizon the proof was given by Huisken and Ilmanen in \cite{HuiskenIlmanen}
using inverse mean curvature flow, and for a multiple component horizon the proof was given by Bray in \cite{Bray} using the conformal flow of metrics.
However for general $k$, the inequality \eqref{0.3} remains an important open problem.

In this paper we aim to study a slightly different inequality. Suppose that the initial data have boundary $\partial M$, consisting of an outermost apparent horizon. Each component of the outermost apparent horizon is either a future or past horizon, however it will be assumed that these two types of components do not intersect.
Then a version of the Penrose conjecture asserts that
\begin{gather}\label{0.4}
M_{ADM}\geq\sqrt{\frac{A_{min}}{16\pi}},
\end{gather}
where $A_{min}$ is the minimum area required to enclose $\partial M$. Moreover the rigidity statement is as follows. Equality holds in \eqref{0.4} if and only if $(M,g,k)$
arises from a spacelike slice of the Schwarzschild spacetime, with outerminimizing boundary.


In \cite{BrayKhuri,BrayKhuri1} Bray and the second author proposed a generalized version of the Jang equation \cite{SchoenYau},
designed specifically for application to the Penrose conjecture. This consists of searching for a graph $\Sigma=\{t=f(x)\}$ within
the warped product manifold $(M\times\mathbb{R},g+\phi^{2}dt^{2})$.
The function $\phi\geq 0$ on $M$ is to be chosen naturally depending on the strategy of the proof, and should reduce to the Schwarzschild warping factor in the case of equality.
In \cite{BrayKhuri,BrayKhuri1} the strategy was based on inverse mean curvature flow (IMCF), and although the resulting $\phi$ had some desirable properties it also has the potential to vanish on the interior of $M$, rendering the coupling of the generalized Jang equation to IMCF degenerate. The purpose of the current paper is to utilize an alternate strategy, namely the conformal flow of metrics \cite{Bray}, and find a new choice for $\phi$ (see \eqref{41} below) which is strictly positive away from $\partial M$ and hence leads to a nondegenerate coupling with the generalized Jang equation.

The generalized Jang equation is motivated by the need for they hypersurface $\Sigma$ to
have weakly nonnegative scalar curvature. As in the classical case \cite{SchoenYau}
this may be achieved by solving
\begin{gather}\label{0.5}
H_{\Sigma}-Tr_{\Sigma}K=0,
\end{gather}
where $H_{\Sigma}$ denotes mean curvature and $K$ is a symmetric 2-tensor on $M\times\mathbb{R}$ which is an extension of the initial data $k$. The precise definition
of $K$ is given in \cite{BrayKhuri,BrayKhuri1} along with the following formula for the scalar curvature of $\Sigma$
\begin{gather}\label{0.7}
\overline{R}=16\pi(\mu-J(w))+
|h-K|_{\Sigma}|^{2}+2|q|^{2}
-2\phi^{-1}\overline{\operatorname{div}}(\phi q),
\end{gather}
where $h$ is the second fundamental form of $\Sigma$, $\overline{\operatorname{div}}$
is the divergence operation with respect to the induced metric $\overline{g}=g+\phi^{2}df^{2}$, and $q$ and
$w$ are the 1-forms
\begin{gather}\label{0.8}
w_{i}=\frac{\phi f_{i}}{\sqrt{1+\phi^{2}|\nabla f|^{2}}},\text{
}\text{ }\text{ }\text{ }\text{ }
q_{i}=\frac{\phi f^{j}}{\sqrt{1+\phi^{2}|\nabla f|^{2}}}(h_{ij}-(K|_{\Sigma})_{ij}).
\end{gather}
Observe that the dominant energy condition implies that the 
right-hand side of \eqref{0.7} is manifestly nonnegative except for the divergence term.
In this way, we may view $\overline{R}$ as being `weakly' nonnegative in that 
multiplying by $\phi$ and integrating by parts produces a nonnegative quantity
modulo boundary integrals.

The expression for the generalized Jang equation \eqref{0.5} in local coordinates
is given by
\begin{gather}\label{0.9}
\left(g^{ij}-\frac{\phi^{2}f^{i}f^{j}}{1+\phi^{2}|\nabla f|^{2}}\right)
\left(\frac{\phi\nabla_{ij}f+\phi_{i}f_{j}+\phi_{j}f_{i}}
{\sqrt{1+\phi^{2}|\nabla f|^{2}}}-k_{ij}\right) = 0.
\end{gather}
Note that this quasilinear elliptic equation is degenerate where $f$ blows-up or where $\phi=0$. Moreover, this equation reduces to the classical Jang equation utilized by Schoen and Yau in their proof of the positive mass theorem \cite{SchoenYau}, when $\phi=1$.

In \cite{HanKhuri} blow-up solutions of the generalized Jang equation have been studied in detail, and the main result may be described as follows. Let $\tau(x)=\operatorname{dist}(x,\partial M)$ and denote the corresponding level sets by
$S_{\tau}$. According to the assumptions above, the boundary may be decomposed into future ($+$) and past ($-$) apparent horizon components $\partial M=\partial_{+}M\cup\partial_{-}M$.
We then specify that in a neighborhood of $\partial_{\pm} M$ there are constants $l\geq 1$ and $c>0$ such that
\begin{gather}
|\theta_{\pm}(S_{\tau})|\leq c\tau^{l}.
\end{gather}
Suppose also that $\phi$ is positive on the interior of $M$, and near $\partial M$ the following structure condition holds
\begin{gather}
\phi(x)=\tau^{b}(x)\widetilde{\phi}(x),
\end{gather}
where the constant $b$ is nonnegative and $\widetilde{\phi}>0$ is a smooth function up to the boundary.
If $\frac{1}{2}\leq b<\frac{l+1}{2}$,
then there exists a smooth solution $f$ of the generalized
Jang equation \eqref{0.9} with the property that
$f(x)\rightarrow\pm\infty$ as $x\rightarrow\partial_{\pm}M$. Furthermore, the asymptotics for this blow-up are given by
\begin{align}\label{0.12}
\begin{split}
\alpha^{-1}\tau^{1-2b}+\beta^{-1} &\leq \pm f
\leq \alpha\tau^{1-2b}+\beta\quad\quad\textit{ if }\quad\quad
\frac{1}{2}< b<\frac{l+1}{2},\\
-\alpha^{-1}\log\tau+\beta^{-1} &\leq \pm f
\leq -\alpha\log\tau+\beta\quad\quad\textit{ if }\quad\quad b=\frac{1}{2},
\end{split}
\end{align}
for some positive constants $\alpha$, $\beta$. If in addition the warping factor satisfies the following asymptotics at spatial infinity
\begin{gather}
\phi(x)=1+\frac{C}{|x|}+O\left(\frac{1}{|x|^{2}}\right)
\quad\quad\text{ as }\quad\quad |x|\rightarrow\infty
\end{gather}
where $C$ is a constant, then
\begin{gather}\label{0.14}
|\nabla^{m}f|(x)=O(|x|^{-\frac{1}{2}-m})\quad\quad\text{
as }\quad\quad |x|\rightarrow\infty,\quad\quad\quad m=0,1,2.
\end{gather}
This fall-off for the solution guarantees that the ADM energy of the Jang surface and that of the initial data are the same.
Furthermore, the blow-up \eqref{0.12} implies that the Jang surface $\Sigma$ is a manifold with boundary,
when $\frac{1}{2}\leq b<1$. This is in contrast to the blow-up solutions of the classical Jang equation
which approximate an infinitely long cylinder over the horizon. Moreover, simple computations shows that
$\partial\Sigma$ is a minimal surface, and that the area of the boundaries of the Jang surface and initial
data agree $|\partial\Sigma|=|\partial M|$.

It is not always the case that blow-up is the correct boundary behavior for application to the Penrose inequality. Moreover instead of working with $M$,
it will become apparent that we should focus on $M_{0}$, where $M_{0}\subset M$ is the region outside of the outermost minimal area enclosure of $\partial M$.
Assuming that the warping factor $\phi$ vanishes at $\partial M_{0}$ and is positive on the interior of $M_{0}$,
the correct boundary condition for the generalized Jang equation is
\begin{equation}\label{0.15}
\overline{H}_{\partial\Sigma_{0}}=0
\end{equation}
where $\Sigma_{0}$ represents the Jang surface over $M_{0}$, and $\overline{H}_{\partial\Sigma_{0}}$ is the mean curvature of $\partial\Sigma_{0}$. It is conjectured \cite{BrayKhuri1} that the
following Neumann type condition at $\partial M_{0}$ implies that \eqref{0.15} holds:
\begin{equation}\label{0.16}
\frac{\phi\partial_{\nu}f(x)}{\sqrt{1+\phi^{2}|\nabla f|^{2}}}=\begin{cases}
+1 & \text{if $\theta_{+}(x)=0$ and $\theta_{-}(x)\neq 0$},\\
-1 & \text{if $\theta_{+}(x)\neq 0$ and $\theta_{-}(x)=0$},\\
0  &  \text{if $\theta_{+}(x)=0$ and $\theta_{-}(x)=0$},
\end{cases}
\end{equation}
where $\nu$ denotes the unit outer normal to $\partial M_{0}$ (pointing away from spatial infinity). Notice that since $\phi|_{\partial M_{0}}=0$
we still have $|\partial\Sigma_{0}|=|\partial M_{0}|$, even if blow-up does not occur.

These results show that the Jang surface $(\Sigma,\overline{g})$ may be interpreted
as a deformation of the initial data $(M,g,k)$, which preserves the geometric quantities involved in the Penrose
inequality, and yields weakly nonnegative scalar curvature. All of this suggests that in order to establish \eqref{0.4},
one should try to apply the techniques used to prove the Riemannian Penrose inequality, inside the Jang surface.
In fact such a method was outlined in \cite{BrayKhuri} and \cite{BrayKhuri1}, for the inverse mean curvature flow.
This leads to a coupling of the inverse mean curvature flow with the generalized Jang equation, through a specific
choice of warping factor $\phi$. Unfortunately, this choice of warping factor lacks regularity and vanishes identically when the weak
inverse mean curvature flow jumps, which in turn leads to a very degenerate generalized Jang equation. Thus, it is of
significant interest to find a more appropriate choice for $\phi$ that leads to a better system of equations. In \cite{BrayKhuri,BrayKhuri1}
it was conjectured that such a $\phi$ exists, and may be found by coupling the generalized Jang equation to the
conformal flow of metrics. It is the purpose of this paper to confirm this, and to give an explicit and simple expression for
$\phi$ (see \eqref{41}). This choice of warping factor is positive on the interior of $M$, and
has better regularity properties than that arising from the inverse mean curvature flow. It will also be shown that
\eqref{41} reduces to the correct expression, that is, the warping factor for the Schwarzschild spacetime, in the
case of equality in \eqref{0.4}.

\begin{theorem}
The Penrose conjecture \eqref{0.4}, along with its corresponding rigidity statement, reduce to the problem of solving a system of PDE arising from a natural coupling of the
(nondegenerate) generalized Jang equation and the conformal flow of metrics.
\end{theorem}

Beyond the Penrose inequality, the generalized Jang equation has been applied to other related geometric inequalities in \cite{DisconziKhuri,ChaKhuriSakovich,Khuri1,Khuri2,KhuriWeinstein}. Moreover in \cite{ChaKhuri1,ChaKhuri2}, a further modification of the Jang equation has been adapted to treat inequalities for which the model spacetime is stationary. Lastly, a charged version of the conformal flow has been developed in \cite{KhuriWeinsteinYamada,KhuriWeinsteinYamada1,KhuriWeinsteinYamada2}, and was used to establish the Penrose inequality with charge for multiple
black holes, in the time-symmetric case. It is likely that the methods of the current note may be
extended to this setting, yielding a coupling of the charged conformal flow with the generalized Jang equation, which may be applied to the Penrose inequality with charge in the non-time-symmetric case.

\section{The Conformal Flow of Metrics and a Refined Version of the Riemannian Penrose Inequality}
\label{sec2} \setcounter{equation}{0}
\setcounter{section}{2}

The purpose of this section is to review the basic properties of the conformal flow of metrics \cite{Bray}, and to obtain a strengthened version
of the Riemannian inequality. We will also slightly modify a portion of Bray's proof of the Riemannian Penrose inequality, so as to make the conformal flow applicable to the setting described in the introduction.

Let $(M,g)$ be a 3-dimensional Riemannian manifold, which is asymptotically flat (having one end), and has
an outerminimizing minimal surface boundary consisting of a finite number of components. Unlike \cite{Bray}, however, we do \textit{not}
assume that the scalar curvature $R_{g}$ is nonnegative. The conformal flow of metrics is given by $g_{t}=u_{t}^{4}g$ where
\begin{equation}\label{1}
\frac{d}{dt}u_{t}=v_{t}u_{t},
\end{equation}
and
\begin{equation}\label{2}
\Delta_{g_{t}}v_{t}=0,\text{ }\text{ }\text{ on }\text{ }\text{ }M_{t},
\end{equation}
\begin{equation}\label{3}
v_{t}|_{\partial M_{t}}=0,\text{ }\text{ }\text{ }v_{t}(x)\rightarrow -1\text{ }\text{ }\text{ as }\text{ }\text{ }
|x|\rightarrow\infty.
\end{equation}
Here $M_{t}$ denotes the region outside of the outermost minimal surface (denoted $\partial M_{t}$) in $(M,g_{t})$.
Note that $u_{t}(x)=\exp\left(\int_{0}^{t}v_{s}(x)ds\right)$. Moreover if $L_{g}$ denotes the conformal Laplacian, then by a standard formula
\begin{equation}\label{4}
u_{t}^{5}R_{g_{t}}=-8L_{g}u_{t}:=-8\left(\Delta_{g}u_{t}-\frac{1}{8}R_{g}u_{t}\right),
\end{equation}
so that
\begin{align}\label{5}
\begin{split}
\frac{d}{dt}(u_{t}^{5}R_{g_{t}})&=-8L_{g}\frac{d}{dt}u_{t}\\
&=-8L_{g}(v_{t}u_{t})\\
&=-8u_{t}^{5}L_{g_{t}}v_{t}\\
&=-8u_{t}^{5}\left(\Delta_{g_{t}}v_{t}-\frac{1}{8}R_{g_{t}}v_{t}\right)\\
&=(u_{t}^{5}R_{g_{t}})v_{t}.
\end{split}
\end{align}
We then have
\begin{equation}\label{6}
u_{t}^{5}R_{g_{t}}=\exp\left(\int_{0}^{t}v_{s}ds\right)R_{g}=u_{t}R_{g}.
\end{equation}

Let $\widetilde{M}_{t}=M_{t}^{-}\cup M_{t}^{+}$ denote the doubled manifold, reflected across the minimal surface
$\partial M_{t}$. Here $M_{t}^{-}$ and $M_{t}^{+}$ each represent a copy of $M_{t}$, and we endow each of these copies
with the metric $g_{t}^{\pm}=(w_{t}^{\pm})^{4}g_{t}$ where $w_{t}^{\pm}=(1\pm v_{t})/2$. The metric on $\widetilde{M}_{t}$ will be denoted by
$\widetilde{g}_{t}=g_{t}^{-}\cup g_{t}^{+}$. Let $c:Cl(T\widetilde{M}_{t})\rightarrow\mathrm{End}(\mathfrak{S})$ be the usual
representation of the Clifford algebra on the bundle of spinors $\mathfrak{S}$, so that
\begin{equation}\label{7}
c(X)c(Y)+c(Y)c(X)=-2g(X,Y).
\end{equation}
Choose orthonormal frame fields $e_{i}$, $i=1,2,3$ for $g$, and observe that $e_{i}^{\pm}=(w_{t}^{\pm}u_{t})^{-2}e_{i}$, $i=1,2,3$ are the corresponding orthonormal
frames for $g_{t}^{\pm}$. There exists a positive definite inner product on $\mathfrak{S}$, denoted by $\langle\cdot,\cdot\rangle$, with respect to which $c(e_{i})$ is anti-hermitian
\begin{equation}\label{8}
\langle c(e_{i})\psi,\eta\rangle=-\langle \psi,c(e_{i})\eta\rangle.
\end{equation}
Define the respective spin connections by
\begin{equation}\label{10}
\nabla_{e_{i}}=e_{i}+\frac{1}{4}\sum_{j,l=1}^{3}\Gamma_{ij}^{l}c(e_{j})c(e_{l}),\text{ }\text{ }\text{ }\text{ }
\nabla_{e_{i}^{\pm}}^{\pm}=e_{i}^{\pm}+\frac{1}{4}\sum_{j,l=1}^{3}\Gamma_{ij}^{\pm l}c(e_{j}^{\pm})c(e_{l}^{\pm}),
\end{equation}
where $\Gamma_{ij}^{l}$ and $\Gamma_{ij}^{\pm l}$ are Levi-Civita connection coefficients for $g$ and $g_{t}^{\pm}$. The corresponding Dirac operators are given by
\begin{equation}\label{11}
\mathcal{D}=\sum_{i=1}^{3}c(e_{i})\nabla_{e_{i}},\text{
}\text{ }\text{ }\text{ }\mathcal{D}^{\pm}=\sum_{i=1}^{3}c(e_{i}^{\pm})\nabla_{e_{i}^{\pm}}^{\pm}.
\end{equation}

Let $\widetilde{\psi}_{t}$ be a harmonic spinor on $\widetilde{M}_{t}$ which converges to a constant spinor of unit norm at spatial infinity; for the present discussion it is assumed that such a spinor exists. We will denote the restriction of
$\widetilde{\psi}_{t}$ to $M_{t}^{\pm}$ by $\psi_{t}^{\pm}$, then $\psi_{t}^{\pm}$ is a harmonic spinor on $(M_{t}^{\pm},g_{t}^{\pm})$.
We may now apply the Lichnerowicz formula \cite{BartnikChrusciel} on each part of the doubled manifold to find that
\begin{equation}\label{12}
 4\pi \widetilde{E}_{ADM}(t)
-\int_{\partial M_{t}^{-}}\langle\psi_{t}^{-},c(e_{3}^{-})\sum_{i=1}^{2}c(e_{i}^{-})\nabla_{e_{i}^{-}}\psi_{t}^{-}\rangle
=\int_{M_{t}^{-}}\left(|\nabla^{-}\psi_{t}^{-}|^{2}+\frac{1}{4}R_{g_{t}^{-}}|\psi_{t}^{-}|^{2}\right)d\omega_{g_{t}^{-}},
\end{equation}
and
\begin{equation}\label{13}
-\int_{\partial M_{t}^{+}}\langle\psi_{t}^{+},c(e_{3}^{+})\sum_{i=1}^{2}c(e_{i}^{+})\nabla_{e_{i}^{+}}\psi_{t}^{+}\rangle
=\int_{M_{t}^{+}}\left(|\nabla^{+}\psi_{t}^{+}|^{2}+\frac{1}{4}R_{g_{t}^{+}}|\psi_{t}^{+}|^{2}\right)d\omega_{g_{t}^{+}},
\end{equation}
where $e_{i}$, $i=1,2$ are tangent and $e_{3}$ is normal (pointing towards spatial infinity) to the boundary $\partial M_{t}$, and $\widetilde{E}_{ADM}(t)$ is the ADM energy of
$(\widetilde{M}_{t},\widetilde{g}_{t})$. A standard calculation shows that
\begin{equation}\label{14}
c(e_{3}^{\pm})\sum_{i=1}^{2}c(e_{i}^{\pm})\nabla_{e_{i}^{\pm}}\psi_{t}^{\pm}=\mathcal{D}_{\partial M_{t}^{\pm}}\psi_{t}^{\pm}-\frac{1}{2}H_{t}^{\pm}\psi_{t}^{\pm},
\end{equation}
where $H_{t}^{\pm}$ are the mean curvatures of $\partial M_{t}^{\pm}$ with respect to the metrics $g_{t}^{\pm}$, and the boundary Dirac operator is given by
\begin{equation}\label{15}
\mathcal{D}_{\partial M_{t}^{\pm}}=c(e_{3}^{\pm})\sum_{i=1}^{2}c(e_{i}^{\pm})\widetilde{\nabla}_{e_{i}^{\pm}}^{\pm}
\end{equation}
with
\begin{equation}\label{16}
\widetilde{\nabla}_{e_{i}^{\pm}}^{\pm}=e_{i}^{\pm}+\frac{1}{4}\sum_{j,l=1}^{2}\Gamma_{ij}^{\pm l}
c(e_{j}^{\pm})c(e_{l}^{\pm}),\text{ }\text{ }\text{ }\text{ }i=1,2.
\end{equation}
We claim that the sum of the boundary terms in \eqref{12} and \eqref{13} all cancel. To see this note that $\psi_{t}^{-}|_{\partial M_{t}^{-}}
=\psi_{t}^{+}|_{\partial M_{t}^{+}}$, and since $v_{t}|_{\partial M_{t}}=0$ we have $g_{t}^{-}|_{\partial M_{t}^{-}}=g_{t}^{+}|_{\partial M_{t}^{+}}$.
Moreover, as $\partial M_{t}^{-}$ and $\partial M_{t}^{+}$ represent the same surface in $\widetilde{M}_{t}$ we have that $e_{3}^{-}=-e_{3}^{+}$, and thus
\begin{equation}\label{18}
\mathcal{D}_{\partial M_{t}^{-}}=-\mathcal{D}_{\partial M_{t}^{+}}.
\end{equation}
Now consider the mean curvature terms. According to a standard formula for the change of mean curvature under conformal deformation
\begin{equation}\label{19}
H_{t}^{\pm}=(w_{t}^{\pm})^{-2}[H_{t}+4u_{t}^{-2}e_{3}(\log w_{t}^{\pm})]=\pm 4u_{t}^{-2}e_{3}(v_{t}),
\end{equation}
where $H_{t}=0$ is the mean curvature of $\partial M_{t}$ with respect to $g_{t}$. It follows that
\begin{equation}\label{20}
H_{t}^{-}=-H_{t}^{+}.
\end{equation}
We may now add \eqref{12} and \eqref{13}, and apply \eqref{18} and \eqref{20}, to obtain
\begin{eqnarray}\label{21}
4\pi\widetilde{E}_{ADM}(t)
&=&\int_{M_{t}^{-}}\left(|\nabla^{-}\psi_{t}^{-}|^{2}+\frac{1}{4}R_{g_{t}^{-}}|\psi_{t}^{-}|^{2}\right)d\omega_{g_{t}^{-}}\\
& &+\int_{M_{t}^{+}}\left(|\nabla^{+}\psi_{t}^{+}|^{2}+\frac{1}{4}R_{g_{t}^{+}}|\psi_{t}^{+}|^{2}\right)d\omega_{g_{t}^{+}}.\nonumber
\end{eqnarray}
Moreover, since $v_{t}$ is harmonic we find
\begin{equation}\label{22}
R_{g_{t}^{\pm}}=-8(w_{t}^{\pm})^{-5}\left(\Delta_{g_{t}}\left(\frac{1\pm v_{t}}{2}\right)-\frac{1}{8}R_{g_{t}}\left(\frac{1\pm v_{t}}{2}\right)\right)=(w_{t}^{\pm}u_{t})^{-4}R_{g}.
\end{equation}
It follows that
\begin{align}\label{24}
\begin{split}
4\pi\widetilde{E}_{ADM}(t)=&\int_{M_{t}}[(w_{t}^{-}u_{t})^{6}|\nabla^{-}\psi_{t}^{-}|^{2}+(w_{t}^{+}u_{t})^{6}|\nabla^{+}\psi_{t}^{+}|^{2}]d\omega_{g}\\
& +\int_{M_{t}}\frac{1}{4}((w_{t}^{-})^{2}|\psi_{t}^{-}|^{2}+(w_{t}^{+})^{2}|\psi_{t}^{+}|^{2})u_{t}^{2}R_{g}d\omega_{g},
\end{split}
\end{align}

It should be noted that a formula similar to \eqref{24} may be obtained without the use of spinors. To see this, solve the zero scalar curvature equation
\begin{equation}\label{24.1}
\Delta_{\widetilde{g}_{t}}z_{t}-\frac{1}{8}\widetilde{R}_{t}z_{t}=0\text{ }\text{ }\text{ on }\text{ }\text{ }\widetilde{M}_{t},\text{ }\text{ }\text{ }\text{ }z_{t}\rightarrow 1\text{ }\text{ }
\text{ as }\text{ }\text{ }|x|\rightarrow\infty,
\end{equation}
where $\widetilde{R}_{t}$ is the scalar curvature of $\widetilde{g}_{t}$; for the present discussion it is assumed that such a solution exists.
Then the conformally related metric $z_{t}^{4}\widetilde{g}_{t}$ has zero scalar curvature. It follows that the ADM energy
is nonnegative $E_{ADM}(z_{t}^{4}\widetilde{g}_{t})\geq 0$. Moreover, the solution has an asymptotic expansion of the form
\begin{equation}\label{24.2}
z_{t}=1+\frac{C_{t}}{|x|}+O\left(\frac{1}{|x|^{2}}\right)\text{ }\text{ }\text{ as }\text{ }\text{ }|x|\rightarrow\infty.
\end{equation}
A small calculation then shows that
\begin{equation}\label{24.3}
0\leq E_{ADM}(z_{t}^{4}\widetilde{g}_{t})=\widetilde{E}_{ADM}(t)+2C_{t}.
\end{equation}
Hence, integrating by parts produces
\begin{align}\label{24.4}
\begin{split}
2\pi\widetilde{E}_{ADM}(t)\geq &-4\pi C_{t}\\
=&\int_{S_{\infty}}z_{t}\partial_{\nu}z_{t}\\
=&\int_{\widetilde{M}_{t}}\left(|\nabla_{\widetilde{g}_{t}}z_{t}|^{2}+\frac{1}{8}\widetilde{R}_{t}z_{t}^{2}\right)d\omega_{\widetilde{g}_{t}}\\
=&\int_{M_{t}}\left((w_{t}^{-}u_{t})^{6}|\nabla_{\widetilde{g}_{t}}z_{t}^{-}|^{2}+(w_{t}^{+}u_{t})^{6}|\nabla_{\widetilde{g}_{t}}z_{t}^{+}|^{2}\right)d\omega_{g}\\
&+\int_{M_{t}}\frac{1}{8}\left[(w_{t}^{-}z_{t}^{-})^{2}+(w_{t}^{+}z_{t}^{+})^{2}\right]u_{t}^{2}R_{g}d\omega_{g},
\end{split}
\end{align}
where $z_{t}^{\pm}$ denotes $z_{t}$ restricted to $M_{t}^{\pm}$. This formula is analogous to \eqref{24}, and was obtained without the use of spinors.
Note that part of the integrand on the
right-hand side of \eqref{24.4} satisfies an elliptic equation with respect to the metric $g$, even though the solution depends on $t$:
\begin{equation}\label{24.5}
0=(w_{t}^{\pm}u_{t})^{5}L_{\widetilde{g}_{t}^{\pm}}z_{t}^{\pm}=L_{g}(w_{t}^{\pm}z_{t}^{\pm}u_{t})
\end{equation}
where $\widetilde{g}_{t}^{\pm}=(z_{t}^{\pm})^{4}g_{t}^{\pm}$.

We mention that an expression similar to \eqref{24} and \eqref{24.4} may also be obtained using the inverse mean curvature flow starting from a point in $\widetilde{M}_{t}$.

Equations \eqref{24} and \eqref{24.4} imply that $\widetilde{E}_{ADM}(t)\geq 0$ under the hypothesis of nonnegative scalar curvature $R_{g}\geq 0$. Actually, this fact may be considered
the crux of the argument for Bray's proof of the Riemannian Penrose inequality. Although the explicit expression above for the energy $\widetilde{E}_{ADM}(t)$ is not necessary for
Bray's proof, it will however be needed in the next section where we outline an argument for the general Penrose inequality. The purpose of \eqref{24} and \eqref{24.4} is to show that the
ADM energy of $(M_{t},g_{t})$, which will be denoted by $m_{t}$, is decreasing when $R_{g}\geq 0$. For the Riemannian Penrose inequality $m_{t}$ is decreasing instantaneously, that is the
derivative $\frac{dm_{t}}{dt}\leq 0$ for all $t$. However for the general Penrose inequality, where one does not necessarily have nonnegative scalar curvature, $m_{t}$ will only be shown to
be decreasing on the interval $[0,\infty)$, that is $m_{0}\geq \lim_{t\rightarrow\infty}m_{t}$. In order to obtain such a result, let us first conclude this review of the proof of the Riemannian
Penrose inequality.

As in \cite{Bray}, we may assume without loss of generality, that the initial data metric $g$ is harmonically flat at
infinity. This means that outside a large compact set, $g$ is scalar flat and conformal to the Euclidean metric, with a conformal factor that approaches a positive constant at infinity.
In this outer region we may then write
\begin{equation}\label{26}
g=\left(1+\frac{m}{r}+O\left(\frac{1}{r^{2}}\right)\right)^{4}\delta.
\end{equation}
Recall that according to a standard formula, if $w=1+\frac{c}{r}+\cdots$ at spatial infinity, then $E_{ADM}(w^{4}g)=E_{ADM}(g)+2c$. Therefore the parameter $m$ in \eqref{26} represents
half the mass of $g$, or rather $m_{0}=2m$. Since $v_{t}$ is a harmonic function, both $u_{t}$ and $v_{t}$ have similar asymptotic expansions at spatial infinity
\begin{equation}\label{27}
u_{t}=\alpha_{t}+\frac{\beta_{t}}{r}+O\left(\frac{1}{r^{2}}\right),\text{ }\text{ }\text{ }\text{ }v_{t}=-1+\frac{\gamma_{t}}{r}+O\left(\frac{1}{r^{2}}\right),
\end{equation}
where $\alpha_{0}=1$ and $\beta_{0}=0$. It follows that
\begin{equation}\label{28}
g_{t}=u_{t}^{4}g=\left(\alpha_{t}+\frac{\beta_{t}+m\alpha_{t}}{r}+O\left(\frac{1}{r^{2}}\right)\right)^{4}\delta,
\end{equation}
and hence
\begin{equation}\label{29}
m_{t}=2\alpha_{t}(\beta_{t}+m\alpha_{t}).
\end{equation}
In order to calculate the derivative $m_{t}'=\frac{dm_{t}}{dt}$, observe that \eqref{1} and \eqref{27} imply
\begin{equation}\label{30}
\alpha_{t}'=-\alpha_{t},\text{ }\text{ }\text{ }\text{ }\beta_{t}'=\alpha_{t}\gamma_{t}-\beta_{t}\text{ }\text{ }\text{ }\text{ }
\Rightarrow \text{ }\text{ }\text{ }\text{ }\alpha_{t}=e^{-t},\text{ }\text{ }\text{ }\text{ }\beta_{t}=e^{-t}\int_{0}^{t}\gamma_{s}ds.
\end{equation}
From this we obtain
\begin{equation}\label{31}
m_{t}'=2\alpha_{t}'(\beta_{t}+m\alpha_{t})+2\alpha_{t}(\beta_{t}'+m\alpha_{t}')=2e^{-2t}\gamma_{t}-4e^{-t}(\beta_{t}+e^{-t}m)=-2(m_{t}-e^{-2t}\gamma_{t}).
\end{equation}
However if we set $\widetilde{g}_{t}=\left(\frac{1-v_{t}}{2}\right)^{4}g_{t}$ then
\begin{equation}\label{32}
\widetilde{g}_{t}=\left(\alpha_{t}+\frac{\beta_{t}+m\alpha_{t}-\alpha_{t}\gamma_{t}/2}{r}+O\left(\frac{1}{r^{2}}\right)\right)^{4}\delta,
\end{equation}
and thus the energy of $\widetilde{g}_{t}$ is given by
\begin{equation}\label{33}
\widetilde{m}_{t}=2\alpha_{t}\left(\beta_{t}+m\alpha_{t}-\frac{\alpha_{t}\gamma_{t}}{2}\right)=m_{t}-e^{-2t}\gamma_{t},
\end{equation}
which yields
\begin{equation}\label{34}
m_{t}'=-2\widetilde{m}_{t}.
\end{equation}
Therefore upon integrating and rewriting in terms of the previous notation, we find that
\begin{equation}\label{35}
E_{ADM}(0)-E_{ADM}(\infty)=2\int_{0}^{\infty}\widetilde{E}_{ADM}(t)dt
\end{equation}
where $E_{ADM}(t)$ denotes the total energy of $(M_{t},g_{t})$.

Together, \eqref{24} (or \eqref{24.4}) and \eqref{35} show that the function $E_{ADM}(t)$ is decreasing over all time. There are two other crucial properties of the conformal
flow which yield the Penrose inequality. First, due to the boundary condition \eqref{3} and the fact that $\partial M_{t}$ is a minimal surface, it follows that
the area function $A(t)$ remains constant in time, where $A(t)$ denotes the area of $\partial M_{t}$. Second, it is shown in \cite{Bray} that after rescaling
$(M_{t},g_{t})$ converges to the exterior region of a constant time slice of the Schwarzschild spacetime, so that $E_{ADM}(\infty)=\sqrt{\frac{A(\infty)}{16\pi}}$.
By combining all these results together we obtain the following refined version of the Penrose inequality
\begin{equation}\label{36}
E_{ADM}-\sqrt{\frac{|\partial M|}{16\pi}}=2\int_{0}^{\infty}\widetilde{E}_{ADM}(t)dt.
\end{equation}
In the case of equality, when $E_{ADM}=\sqrt{\frac{|\partial M|}{16\pi}}$, standard arguments combined with \eqref{24} (or \eqref{24.4}) and \eqref{35} show that $(M,g)$ is conformally
equivalent to $(\mathbb{R}^{3}-B_{1}(0),\delta)$ with zero scalar curvature. Since the boundary $\partial M$ is a minimal surface, it follows that $(M,g)$ is isometric
to a constant time slice of the Schwarzschild spacetime.

\section{Coupling the Conformal Flow to the Generalized Jang Equation}
\label{sec3} \setcounter{equation}{0}
\setcounter{section}{3}


Let $\Sigma_{0}$ be a solution of the generalized Jang equation \eqref{0.9} satisfying boundary condition \eqref{0.15}, as described in Section \ref{sec1}. Then the boundary
$\partial\Sigma_{0}$ is a minimal surface, and is in fact the outermost minimal surface. To see this, suppose that a surface $\overline{S}$ encloses but is not equal to $\partial\Sigma_{0}$ and
satisfies $|\overline{S}|\leq |\partial\Sigma_{0}|$. Since $\overline{g}\geq g$, we have $|S|\leq|\overline{S}|$, where $S\subset M$ is the projection of $\overline{S}$. Thus
$|S|\leq|\partial\Sigma_{0}|=|\partial M_{0}|$, which is impossible since $\partial M_{0}$ is the outermost minimal area enclosure of $\partial M$.

We will now perform the conformal flow of metrics on the Jang surface $\Sigma_{0}$, and show how this leads naturally to a choice for the warping factor $\phi$.
Following the notation from the previous section let $(g_{t}=u_{t}^{4}\overline{g},\Sigma_{t})$ be a conformal flow,
where $\Sigma_{t}$ denotes the region outside of the outermost minimal surface $\partial \Sigma_{t}$. Then from \eqref{6} the scalar curvature evolves by
\begin{equation}\label{37}
R_{g_{t}}=u_{t}^{-4}\overline{R}.
\end{equation}
Moreover, since the ADM energies of $(M,g,k)$ and $(\Sigma,\overline{g})$ agree, we may apply \eqref{24} or \eqref{24.4} and \eqref{35} to obtain
\begin{equation}\label{38}
E_{ADM}-\sqrt{\frac{|\partial\Sigma_{0}|}{16\pi}}\geq 2\int_{0}^{\infty}\widetilde{E}_{ADM}(t)dt,
\end{equation}
where $\widetilde{E}_{ADM}(t)$ denotes the total energy of $(\widetilde{\Sigma}_{t},g_{t}^{-}\cup g_{t}^{+})$ and is given by
\begin{align}\label{39}
\begin{split}
4\pi\widetilde{E}_{ADM}(t)=&\int_{\Sigma_{t}}[(w_{t}^{-}u_{t})^{6}|\nabla^{-}\psi_{t}^{-}|^{2}+(w_{t}^{+}u_{t})^{6}|\nabla^{+}\psi_{t}^{+}|^{2}]d\omega_{\overline{g}}\\
& +\int_{\Sigma_{t}}\frac{1}{4}((w_{t}^{-})^{2}|\psi_{t}^{-}|^{2}+(w_{t}^{+})^{2}|\psi_{t}^{+}|^{2})u_{t}^{2}\overline{R}d\omega_{\overline{g}},
\end{split}
\end{align}
or
\begin{align}\label{39.1}
\begin{split}
4\pi\widetilde{E}_{ADM}(t)
=&\int_{\Sigma_{t}}2\left((w_{t}^{-}u_{t})^{6}|\nabla_{\widetilde{g}_{t}}z_{t}^{-}|^{2}+(w_{t}^{+}u_{t})^{6}|\nabla_{\widetilde{g}_{t}}z_{t}^{+}|^{2}\right)d\omega_{\overline{g}}\\
&+\int_{\Sigma_{t}}\frac{1}{4}\left[(w_{t}^{-}z_{t}^{-})^{2}+(w_{t}^{+}z_{t}^{+})^{2}\right]u_{t}^{2}\overline{R}d\omega_{\overline{g}}.
\end{split}
\end{align}

Notice that we cannot conclude as in Section \ref{sec2} that $\widetilde{E}_{ADM}(t)\geq 0$, since the scalar curvature $\overline{R}$ of the Jang surface is
not necessarily nonnegative. According to \eqref{0.7}, it is a divergence term which is the possible obstruction to nonnegativity. However by choosing $\phi$
appropriately, this divergence term may be integrated away so that the right-hand side of \eqref{36} is nonnegative. To accomplish this let
\begin{equation}\label{40}
\chi_{t}(x)=\begin{cases}
1 & \text{if $x\in\Sigma_{t}$},\\
0 & \text{if $x\in\Sigma_{0}-\Sigma_{t}$},
\end{cases}
\end{equation}
and set
\begin{equation}\label{41}
\phi=2\int_{0}^{\infty}\chi_{t}((w_{t}^{-})^{2}|\psi_{t}^{-}|^{2}+(w_{t}^{+})^{2}|\psi_{t}^{+}|^{2})u_{t}^{2}dt,
\end{equation}
or
\begin{equation}\label{41.1}
\phi=2\int_{0}^{\infty}\chi_{t}\left[(w_{t}^{-}z_{t}^{-})^{2}+(w_{t}^{+}z_{t}^{+})^{2}\right]u_{t}^{2}dt.
\end{equation}
Notice that $\phi$ should be strictly positive away from $\partial\Sigma_{0}$,
\begin{equation}\label{41.2}
\phi(x)\rightarrow 1\text{ }\text{ }\text{ as }\text{ }\text{ }|x|\rightarrow\infty,\text{ }\text{ }\text{ and }\text{ }\text{ }
\phi|_{\partial\Sigma_{0}}=0,
\end{equation}
where we have used \eqref{3} and the fact that $u_{t}\rightarrow e^{-t}$ as $|x|\rightarrow\infty$.
By applying Fubini's Theorem, using the boundary condition \eqref{41.2}, as well as the scalar curvature formula for the Jang surface \eqref{0.7},
we may calculate
\begin{align}\label{42}
\begin{split}
\int_{0}^{\infty}\int_{\Sigma_{t}}((w_{t}^{-})^{2}|\psi_{t}^{-}|^{2}+(w_{t}^{+})^{2}|\psi_{t}^{+}|^{2})u_{t}^{2}\overline{R}d\omega_{\overline{g}}dt
&=\frac{1}{2}\int_{\Sigma_{0}}\phi\overline{R}d\omega_{\overline{g}}\\
&\geq\frac{1}{2}\int_{\Sigma_{0}}\phi\left(2|q|^{2}-\frac{2}{\phi}\overline{div}(\phi q)\right)d\omega_{\overline{g}}\\
&=\int_{\Sigma_{0}}\phi|q|^{2}d\omega_{\overline{g}}
+\int_{\partial\Sigma_{0}}\phi q(\overline{N})d\sigma_{\overline{g}}\\
&=\int_{\Sigma_{0}}\phi|q|^{2}d\omega_{\overline{g}},
\end{split}
\end{align}
where $\overline{N}$ is the unit normal to $\partial\Sigma_{0}$ pointing towards spatial infinity. We could also have used \eqref{41.1}.
The above calculation also uses the reasonable assumption that $|q|$ remains uniformly bounded and decays very fast at spatial infinity; this behavior
is consistent with that of solutions to the classical Jang equation \cite{SchoenYau}.
This may be combined with \eqref{0.7}, \eqref{38}, \eqref{39}, and \eqref{42} to yield
\begin{align}\label{43}
\begin{split}
& E_{ADM}-\sqrt{\frac{|\partial\Sigma_{0}|}{16\pi}}\\
\geq&\frac{1}{2\pi}\int_{0}^{\infty}\int_{\Sigma_{t}}[(w_{t}^{-}u_{t})^{6}|\nabla^{-}\psi_{t}^{-}|^{2}+(w_{t}^{+}u_{t})^{6}|\nabla^{+}\psi_{t}^{+}|^{2}]d\omega_{\overline{g}}dt\\
& +\frac{1}{8\pi}\int_{0}^{\infty}\int_{\Sigma_{t}}((w_{t}^{-})^{2}|\psi_{t}^{-}|^{2}+(w_{t}^{+})^{2}|\psi_{t}^{+}|^{2})u_{t}^{2}
\left(16\pi(\mu-J(w))+|h-K|_{\Sigma}|^2+2|q|^2\right)d\omega_{\overline{g}}.
\end{split}
\end{align}
The Penrose inequality \eqref{0.4} (with ADM energy replacing ADM mass) now follows. This is due to the fact that the Jang metric $\overline{g}$
measures areas to be at least as large as those measure by $g$, and hence $|\partial\Sigma_{0}|\geq A_{min}$. We remark that because of the asymptotics \eqref{0.14}, only the ADM energy is encoded in the solution of the generalized Jang equation. At this time it is unknown how to encode the ADM linear momentum into a solution of Jang's equation.

Now consider the case of equality in \eqref{0.4}. We will use \eqref{41}, although a similar argument holds for \eqref{41.1}. First note that in this case $|\partial\Sigma_{0}|=A_{min}$.
Next, according to \eqref{43}, both spinors $\psi_{t}^{\pm}$ are parallel. This implies that $|\psi_{t}^{\pm}|\equiv 1$, in light of the boundary
condition at spatial infinity. It follows that
\begin{equation}\label{44}
\mu=|J|\equiv 0,\text{ }\text{ }\text{ }\text{ }h=K|_{\Sigma},\text{ }\text{ }\text{ }\text{ }q\equiv 0,
\end{equation}
where we have used the dominant energy condition and the fact that $|w|<1$ away from $\partial M$. Hence $\overline{R}\equiv 0$. Moreover the existence of a basis of parallel spinors shows that
$(\widetilde{\Sigma}_{t},g_{t}^{-}\cup g_{t}^{+})$ is isometric to $(\mathbb{R}^{3},\delta)$. The Jang surface $\Sigma_{0}$ is then conformally flat with zero scalar curvature, and since $\partial\Sigma_{0}$
is minimal, we find that $(\Sigma_{0},\overline{g})$ is isometric to the exterior region of the $t=0$ slice of the Schwarzschild spacetime
\begin{equation}\label{45}
\mathbb{SC}^{4}=(\mathbb{R}\times(\mathbb{R}^{3}-B_{\frac{m_{0}}{2}}(0)),-\phi_{SC}^{2}dt^{2}+g_{SC}),
\end{equation}
where $r=|x|$ and
\begin{equation}\label{46}
\phi_{SC}(x)=\frac{1-\frac{m_{0}}{2r}}{1+\frac{m_{0}}{2r}},\text{ }\text{ }\text{ }\text{ }g_{SC}=\left(1+\frac{m_{0}}{2r}\right)^{4}\delta.
\end{equation}

We will show that $\phi=\phi_{SC}$. Since $\overline{g}$ is isometric to the spatial Schwarzschild metric, the conformal flow of metrics may be given explicitly, namely
\begin{equation}\label{47}
v_{t}(x)= \begin{cases}
            \frac{-e^{-t}+\frac{m_{0}}{2r}e^{t}}{e^{-t}+\frac{m_{0}}{2r}e^{t}} & \quad \text{ if } r \geq \frac{m_{0}}{2}e^{2t} \\
            0 & \quad \text{ if } r < \frac{m_{0}}{2}e^{2t}
        \end{cases},
\text{ }\text{ }\text{ }\text{ }\text{ }\text{ }\text{ }
u_{t}(x)= \begin{cases}
            \frac{e^{-t}+\frac{m_{0}}{2r}e^{t}}{1+\frac{m_{0}}{2r}} & \quad \text{ if } r \geq \frac{m_{0}}{2}e^{2t} \\
            \frac{\sqrt{\frac{2m_{0}}{r}}}{1+\frac{m_{0}}{2r}} & \quad \text{ if } r < \frac{m_{0}}{2}e^{2t}
       \end{cases}.
\end{equation}
It follows that on $\Sigma_{t}$
\begin{equation}\label{48}
w_{t}^{-}=\frac{1-v_{t}}{2}=\frac{e^{-t}}{e^{-t}+\frac{m_{0}}{2r}e^{t}},\text{ }\text{ }\text{ }\text{ }w_{t}^{+}=\frac{1+v_{t}}{2}=\frac{\frac{m_{0}}{2r}e^{t}}{e^{-t}+\frac{m_{0}}{2r}e^{t}},
\end{equation}
and therefore
\begin{align}\label{49}
\begin{split}
\phi(x)=&2\int_{0}^{\log\sqrt{\frac{2r}{m_{0}}}}
\left[\left(\frac{e^{-t}}{e^{-t}+\frac{m_{0}}{2r}e^{t}}\right)^{2}
+\left(\frac{\frac{m_{0}}{2r}e^{t}}{e^{-t}+\frac{m_{0}}{2r}e^{t}}\right)^{2}\right]
\left(\frac{e^{-t}+\frac{m_{0}}{2r}e^{t}}{1+\frac{m_{0}}{2r}}\right)^{2}dt\\
=&2\int_{0}^{\log\sqrt{\frac{2r}{m_{0}}}}\frac{e^{-2t}+\left(\frac{m_{0}}{2r}\right)^{2}
e^{2t}}{\left(1+\frac{m_{0}}{2r}\right)^{2}}dt\\
=&\left(\frac{1-\frac{m_{0}}{2r}}{1+\frac{m_{0}}{2r}}\right)=\phi_{SC}.
\end{split}
\end{align}

Consider now the graph map $G:M\rightarrow\mathbb{SC}^{4}$ given by $x\mapsto(x,f(x))$. Since $\phi\equiv\phi_{SC}$, the map $G$ yields an isometric embedding of the initial data $(M_{0},g,k)$ into the Schwarzschild spacetime, as
\begin{equation}\label{50}
g=\overline{g}-\phi^{2}df^{2}=g_{SC}-\phi_{SC}^{2}df^{2}.
\end{equation}
Moreover since $h=K|_{\Sigma}$, a calculation \cite{BrayKhuri}, \cite{BrayKhuri1} guarantees that the second fundamental form of this embedding agrees with $k$. Lastly observe that since $\phi$ vanishes at $\partial M_{0}$, we must have that $\partial M_{0}$ is an apparent horizon. However $\partial M$ is an outermost apparent horizon, so in fact $\partial M_{0}=\partial M$, and hence $M_{0}=M$.

\begin{remark}
Without further modification, the arguments of this paper can only yield the inequality
\begin{equation}\label{50.1}
E_{ADM}\geq\sqrt{\frac{A_{min}}{16\pi}},
\end{equation}
where the ADM mass is replaced by the ADM energy in \eqref{0.4}. The rigidity statement for \eqref{50.1} is as follows, equality holds only if $(M,g,k)$ arises from a spacelike slice of the Schwarzschild spacetime, with outerminimizing boundary. Note that the `if' direction is not included.

A proof of the `if' part in the rigidity statement for the inequality \eqref{0.4} is as follows. Suppose that $(M,g,k)$ is a spacelike slice of the Schwarzschild spacetime with outerminimizing boundary.
Since $\phi_{SC}|_{\partial M}=0$, we have that $|\partial M|=|\partial\widehat{M}|$, where $(\widehat{M},g_{SC},0)$ is the $t=0$ slice of the Schwarzschild spacetime. Furthermore
\begin{equation}
M_{ADM}=M_{ADM}(\widehat{M})=E_{ADM}(\widehat{M}),
\end{equation}
so that
\begin{equation}\label{50.2}
M_{ADM}=E_{ADM}(\widehat{M})=\sqrt{\frac{|\partial\widehat{M}|}{16\pi}}=\sqrt{\frac{|\partial M|}{16\pi}}\geq\sqrt{\frac{A_{min}}{16\pi}}.
\end{equation}
It follows that equality holds in \eqref{0.4} since $\partial M$ is outerminimizing, that is $|\partial M|=A_{min}$. These same arguments also show why the boundary must be outerminimizing in the case
of equality. Namely, in light of \eqref{50.2}, the only way that equality can hold in \eqref{0.4} is if $|\partial M|=A_{min}$.
\end{remark}

\end{document}